\documentstyle[prl,aps,epsf,multicol,amssymb]{revtex}

\begin{document}
\advance\textheight by 0.5in
\advance\topmargin by -0.25in
\draft

\title{A New Disorder Driven Roughening Transition of CDW's and Flux Line
  Lattices}

\author{Thorsten Emig and Thomas Nattermann}
\address{Institut f\"ur Theoretische Physik, Universit\"at zu K\"oln,
  Z\"ulpicher Str. 77, D-50937 K\"oln, Germany}

\date{\today}

\maketitle

\begin{abstract}
  We study the competition between pinning of a charge density wave
  (CDW) by random distributed impurities and a periodic potential of
  the underlying crystal lattice. In $d=3$ dimensions, we find for
  commensurate phases of order $p>p_c\approx 6/\pi$ a disorder driven
  continuous roughening transition from a 'flat' phase with
  translational long range order to a 'rough' glassy phase with quasi
  long range order. Critical exponents are calculated in a double
  expansion in $\mu=p^2/p_c^2-1$ and $\epsilon=4-d$ and fulfill the
  scaling relations of random field models. Implications for flux line
  lattices in high temperature superconductors are briefly discussed.
\end{abstract}
\begin{multicols}{2}
  
  Charge density waves (CDWs) are broken symmetry states of mostly
  strongly anisotropic quasi-one dimensional metals
  \cite{Fr54,Grbook}.  Their ground state is a coherent superposition
  of electron-hole pairs and shows a periodic spatial modulation with
  wave vector ${\bf q}_0=2{\bf k}_F$.  The Fermi wavevector ${\bf
    k}_F$ enters here due to the divergent electronic response at
  $2{\bf k}_F $ in one dimension.  This in combination with a
  sufficiently large electron-phonon coupling may lead to a Peierls
  instability, resulting in a new state with a periodically varying
  charge modulation of density
\begin{equation}
  \rho({\mathbf x})\approx\rho_0+\rho_1\cos({\bf q}_0{\bf
    x}+\phi({\mathbf x})).
\end{equation}
For a partially filled electron band, the period $\pi/k_F$ is in
general incommensurate with the underlying lattice.  Considering the
excitations of CDWs well below the instability temperature, one may
neglect variations of the amplitude $\rho_1$ since they are
accompanied by relatively high energy expenses. In contrast, long
wavelength deformations of the phase $\phi ({\bf x})$ are accompanied
by arbitrarily small elastic energy costs.  In perfectly clean
crystals the phase excitations form a gapless excitation branch, giving
rise to an infinite conductivity at zero frequency.

A very similar picture is found for other incommensurate structures
like spin density waves in anisotropic metals \cite{Grbook}, mass
density waves in superionic conductors \cite{Bak}, polarization
density waves in incommensurate ferroelectrics \cite{Dvo} or flux-line
arrays in type-II superconductors \cite{Bal95}. In the absence of a
coupling to the underlying lattice, the wavevector ${\bf q}_0$ of the
modulation is related to external parameters like temperature for
ferroelectrics or the external magnetic field for superconductors and
in general incommensurate with one of the reciprocal lattice vectors
${\bf g}$ of the crystal.  Although our considerations apply to this
large group of systems as well, we will continue to use the
terminology of CDWs. Its extension to the other cases is
straightforward and will be discussed for flux-line lattices in
high-temperature superconductors briefly at the end.

Lattice effects, which have been neglected so far, may become
relevant if ${\bf q_0}$ is close to ${\bf g}/p$ where $p$ is an
integer corresponding to $1/p$ filling of the conduction band.  For a
sufficiently strong lattice potential $g^2v$ (in units of the phase
stiffness constant $\gamma $ ), the density wave locks in at the
commensurate wavevector ${\bf g}/p$. In this case phase fluctuations
are suppresed, i.e. the CDW is in a 'flat' phase. The phason branch
acquires a gap of order $v$ and an applied external force density
$f_{\rm ex}$ has to overcome a threshold value $f_{c,v} \sim \gamma
g^2 p v$ to depin the structure.  If ${\bf g} $ and ${\bf q}_0$ are
parallel, the width of the commensurate phase is given by
$|g/p-q_0|=\delta < \delta_c \approx g v^{1/2}$.

The actual modulation of $\phi ({\bf x})$ in these phases is in
general very complicated. It is therefore convenient to go over to a
new phase field which describes the distortions from the lock-in
wavevector, such that $\phi=0$ in the commensurate (C) phase. The
incommensurate (I) phase close to the lock-in is then described by a
lattice of domain walls of intrinsic width $\xi_0 \approx
1/(pgv^{1/2})$ and diverging separation $l$ if the C phase is
approached. There is no lattice pinning in the I phase.

Thermal fluctuations strongly diminish C phases in two
dimensions \cite{Po79}. In three dimensions thermal fluctuations have
only minor effects close to the transition to the disordered phase.
On the other hand, experimentally, commensurability effects have
rarely been seen, which was mainly explained by the smallness of
$v$.

Disorder is another ingredient which is known to change this simple
picture.  If the potential of an individual impurity is weak, it is
sufficient to consider its effect on the phase order only. Neglecting
commensurability effects, it was shown by Fukuyama and Lee \cite{Fu78}
that long range order of phases $\phi({\bf x})$ is indeed destroyed by
an arbitrarily weak random potential in dimensions $d<4$ on length
scales $L$ larger than the Fukuyama-Lee-length
$L_\Delta=\Delta^{-1/(4-d)}$. 
Here $\Delta=-R''(0)$ denotes the strength of the disorder fluctuations, where
$R(\phi)$ is defined in Eq. (\ref{discor}).  As a result of the
adaption of the CDW to the impurities, the modulated structure is now
also pinned in the I phase with a pinning force density
$f_{c,\Delta} \approx \gamma L_\Delta^{-2}$.

In general, however, lattice and impurity pinning will occur
simultaneously, but act in opposite directions. Whereas the lattice
prefers registry of the CDW, resulting in a commensurate {\it flat}
phase, the disorder tends to destroy its long range order and makes
the CDW {\it rough}.  It is therefore the aim of the present paper to
study the {\it combined effect of both pinning mechanisms}. We will
restrict ourselves in the following to the realistic situation where
$gL_\Delta, g\xi_0 \gg 1$, i.e.  to the weak pinning case. Since
pinning forces from the lattice and disorder act in general in
opposite directions with maximal values $f_{c,v}$ and $f_{c,\Delta}$,
respectively, one naively expects that the C phase is stable if
$f_{c,\Delta}<f_{c,v}$, i.e. for $\Delta\lesssim (g^2pv)^{(4-d)/2}$
\cite{Fu78b}.

A more detailed calculation to be presented below however shows that
for $\delta=0$ and $p<p_c(d)$ ($p_c(3)\approx 6/\pi$) the pinning due
to the lattice is dominating even for arbitrarily small potential
values $v$, and the C phase persists up to an intermediate disorder
strength.  Inside the C phase, the translational order parameter
$\Psi({\bf x})=\rho_1\exp(i\phi({\bf x}))$ takes a non-zero
expectation value $\overline{\langle\Psi({\bf x})\rangle}$, where
$\langle\, \rangle$ and the bar denote thermal and configurational
average, respectively.  For $p>p_c(d)$, the system indeed undergoes a
disorder driven continuous roughening transition if $\Delta$ exceeds a
threshold value $\Delta_c\sim v^{\epsilon/2}$ with $\epsilon=4-d$.
For $\Delta<\Delta_c$, $\overline{|\langle\Psi({\bf x})\rangle|} \sim
(\Delta_c-\Delta)^\beta$, and the correlation function $K({\bf
  x})=\overline{\langle\delta\Psi({\bf x})\delta\Psi^\ast({\bf
    0})\rangle}$, $\delta\Psi ({\bf x})=\Psi ({\bf
  x})-\overline{\langle\Psi ({\bf x})\rangle}$, shows an exponential
decay. In contrast, for $\Delta\ge \Delta_c$ the phase is rough and
unlocked despite $\delta=0$ with $\overline{\langle\Psi({\bf
    x})\rangle}=0$ but a power law decay of $K({\bf x})\sim |{\bf
  x}|^{4-d-\bar\eta}$. From a functional RG calculation we find in a
double expansion \cite{ack} in $\epsilon \ll 1$ and $\mu=p^2/p_c^2-1
\ll 1$ the critical exponents
\begin{equation}
\label{exponents}
  \nu^{-1}=4\mu, \quad \frac{\beta}{\nu}=\frac{\pi^2}{18}\epsilon,
  \quad \theta=2-\epsilon,
\end{equation}
where $\nu$ denotes the correlation length exponent in the C phase.
The exponent $\theta$ is the negative temperature eigenvalue which
appears in the random field hyperscaling relation
$\nu(d-\theta)=2-\alpha$ \cite{Vi85}. The critical exponents fulfill
all exponent relations and inequalities known from random field models
with discrete symmetry of the order parameter \cite{Vi85}. In
particular, we obtain the exponent $\bar\eta=\epsilon(1+\pi^2/9)$ at
$\Delta_c$.


Our starting point for the description of the CDW close to a
C phase with a $p$-fold degenerate groundstate is the
following Ginzburg-Landau-Hamiltonian \cite{Grbook}
\begin{equation}
  {\cal H}\!=\!\gamma\!\!\int\! d^d {\mathbf x}\! \left\{
    \frac{1}{2}({\bbox \nabla} \phi - {\bbox \delta})^2 -
    g^2 v \cos(p\phi) + V(\phi,{\mathbf x}) \right\}
\label{Ham}
\end{equation}
with ${\bbox \delta}=\delta\hat{\mathbf z}$.  Here the first term
represents the elastic energy of the long--wavelength phase
deformations controlled by the stiffness constant
$\gamma=n(\epsilon_{\rm F})\hbar^2 v_{\rm F}^2/2$ with Fermi velocity
$v_{\rm F}$ and density of states $n(\epsilon_{\rm F})$ at the Fermi
level. $g^2v\sim \rho_1^{p-2}$ measures the strength of the lattice
potential relative to $\gamma$. The last term $V(\phi,{\bf
  x})=\rho_1V_0 \sqrt{n_i} \cos(\phi({\bf x})-\alpha({\bf x}))$
describes the effect of impurities, where $\gamma V_0$ and $n_i$ are
the strength and the density of impurities, respectively. The
uniformly distributed and uncorrelated random phase $\alpha({\bf x})$
arises from the random impurity positions since we have assumed the
weak pinning case with $n_i L_\Delta^d \gg 1$\cite{Fu78,Fu78b}. The
disorder correlation function $R(\phi)$ is given by
\begin{equation}
\label{discor}
\overline{V(\phi,{\mathbf x})V(\phi',{\mathbf x}')}=R(\phi-\phi') 
\delta({\mathbf x}-{\mathbf x'})
\end{equation}
with a bare function $R_0(\phi)=n_i \rho_1^2 V_0^2 \cos(\phi)/2$.

We first estimate the influence of the pinning potentials by
perturbation theory. For a vanishing lattice potential $v=0$, to begin
with, it has been shown \cite{Vi84} that the phase
correlation function $C({\bf x})=\overline{\langle(\phi({\bf
    x})-\phi({\bf 0}))^2\rangle}$ behaves logarithmically on length
scales $|{\bf x}|>L_\Delta$, i.e.  $C({\bf x}) = 2A(4-d)\ln|{\bf
  x}/L_\Delta|$ with $A=\pi^2/9 \approx 1.1$ \cite{Gi95}.  Neglecting
the non-Gaussian character of $\phi({\bf x})$, which is justified for
$\epsilon\ll 1$, lowest order perturbation theory in $v$ yields an
effective Hamiltonian with mass\cite{subtleties}
\begin{equation}
\label{pert}
  g^2p^2v \overline{\cos p\phi}  \sim e^{-p^2\overline{\phi^2}
  /2} \sim (L/L_\Delta)^{-p^2 A(4-d)/2}
\end{equation}
on length scales $L>L_\Delta$.  Comparing this power of $L$ with the
$L^{-2}$ behavior of the elastic term, we conclude that the
perturbation is always relevant if $p<p_c=(4/A(4-d))^{1/2}$.  For
larger $p$ a small periodic potential can be neglected.  

Next we employ an RG calculation, which starts with the disorder
averaged replica-Hamiltonian
\begin{eqnarray}
{\cal H}_n&=&\gamma\int d^d x \bigg\{\frac{1}{2} \sum_{\alpha} ({\bbox
    \nabla} \phi_\alpha-\delta\hat{\bf z})^2 - g^2v \sum_{\alpha}
  \cos(p\phi_\alpha) \nonumber \\
&& -\frac{\gamma}{2T}\sum_{\alpha\beta}
  R(\phi_\alpha-\phi_\beta)\bigg\}.
\label{RHam}
\end{eqnarray}
In the absence of crystal pinning, $v\equiv 0$, the misfit $\delta$
can be shifted away by the transformation $\phi({\bf x}) \rightarrow
\phi({\bf x})+\delta z$ due to the statistical symmetry of the
disorder under spatial translations. The corresponding Hamiltonian has
been studied repeatedly in the past \cite{Fi86,Gi95} using the
functional RG-technique in $d=4-\epsilon$ dimensions.  The resulting
RG flow leads to a zero-temperature fixed point $U^*$ where
$-R''_U(0)=\Delta^*_U=(\pi/3)^2\epsilon \Lambda^\epsilon K_d^{-1}$,
which describes the glassy unlocked phase (U) of the CDW.

To extend the calculation to $v\neq 0$, we follow here the functional
RG treatment developed by Balents and Kardar for a closely related
case \cite{Bal94}. Although for $p>p_c$ a weak periodic potential is
irrelevant, one expects that it becomes a relevant perturbation once
$v$ exceeds a critical value $v_c$. Since $v_c$ has to vanish if $p
\rightarrow p_c^+$, it is tempting to describe the new fixed point
$F^*$ corresponding to this transition perturbatively in
$\epsilon=4-d$ and $\mu=p^2/p_c^2-1$. A finite value of $v$ close to
$F^*$ will lead in particular to a renormalization of $\gamma/T$.
While keeping $\gamma$ fixed, we absorb this in an additional
renormalization of $T$, $R(\phi)$ and $v$. The fixed point value
$\Delta^*_F$ of $\Delta$ at $F^*$ differs from $\Delta^*_U$ by
corrections of order $\epsilon\mu$. The renormalization of $T$ will
hence not change the sign of $-\theta=-2+{\cal O}(\epsilon,\mu)$,
describing the irrelevance of $T$ also at $F^*$.  
\narrowtext
\begin{figure}[htb]
\begin{center}
\leavevmode
\epsfxsize=1.0\linewidth
\epsfbox{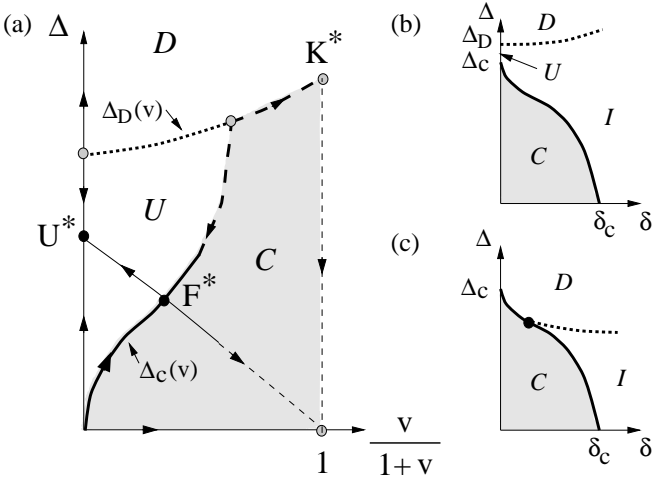}
\end{center}
\caption{(a)  Schematic RG flow for $\delta\equiv 0$ and $p>p_c$ in
  the $v$-$\Delta$-plane. A critical line $\Delta_c(v)$ with fixed
  point $F^*$ separates the flat ({\it C}) and rough, unlocked ({\it
    U}) phase.  For $p<p_c$ the fixed points $F^*$ and $U^*$ merge.
  Since for large disorder strength $\Delta$ topological defects will
  proliferate, we expect a disordered phase {\it D} above the line
  $\Delta_D(v)$ but the RG flow in this range is not yet clear.  The
  transition from the {\it C} to the {\it D} phase is probably in the
  universality class of the random field $p$-state clock model.  (b)
  Possible phase diagrams for finite $\delta$ at fixed $v$ for $p>p_c$
  and (c) for $p<p_c$.}
\label{fig.dia}
\end{figure}
Under the RG transformation higher order harmonics $v_m \cos(pm\phi)$,
$m>1$, will be generated. However, the $v_m$ with $m>1$ are irrelevant
at $F^*$ since the eigenvalues $\lambda_m=2-\frac{1}{2} p^2 m^2
\Delta^*_F = 2(1-m^2(1+{\cal O}(\mu)))$ of these pertubations are
negative for $\mu \rightarrow 0$ if $m>1$. We can restrict ourselves
therefore to the lowest order harmonic $v\cos(p\phi)$.  Note that also
terms of the form $(R''(\phi_\alpha-\phi_\beta))^2
\cos(mp\phi_\alpha)$, generated by the RG, are strongly irrelevant with
scaling dimension $\lambda_m-d-2$.  Collecting the contributions to
the RG-flow up to the first non-linear order in $v$ and $R(\phi)$, we
obtain the following RG flow equations valid close to the $T=0$ fixed
points and to lowest order in $\epsilon$ and $\mu$ with $\Delta=-R''(0)$,
\begin{eqnarray}
 \label{flow0}
 \frac{dT}{dt}&=&\left\{2-d-\frac{p^6}{4}\Delta v^2\right\}T \equiv
 -\theta(v,\Delta) T\\
 \label{flow1}
 \frac{dv}{dt}&=&\left\{2-\frac{p^2}{2}\Delta\right\}v +
 \frac{p^6}{8}\Delta v^3\\
 \label{flow2}
\frac{\partial R(\phi )}{\partial t}&=&\left\{\epsilon-\frac{p^6}{2}
v^2 \Delta \right\} R(\phi)
+ \frac{1}{2}R''(\phi)^2 +\Delta R''(\phi)
\end{eqnarray}
where we made the substitutions $\Lambda^{-2} g^2 v \rightarrow v$,
$K_d\Lambda^{d-4} R(\phi) \rightarrow R(\phi)$,
$K_d^{-1}=2^{d-1}\pi^{d/2}\Gamma(d/2)$. Next we analyze the fixed
points of the RG flow.  Rewriting $R(\phi )$ in a Fourier series
$R(\phi )=\sum_m R_m \cos{m\phi }$, the fixed point function
$R^*_{U}(\phi)$ \cite{Gi95} for $v=0$ (as well for any other fixed
value of $v$) can be parameterized by $\frac{\pi^2}{6} m^4 R_m^{\ast}
=\Delta^{\ast}(v)$.  Rewriting $R_m=\frac{6}{\pi^2}m^{-4}\Delta +
r_m$, the $r_m$ are irrelevant with eigenvalues $-\epsilon(1-\mu)c_m$
with positive factors $c_m\sim m^2$.  Since we are interested in the
universal properties of the transition we will therefore
adopt here the particularly simple
functional form of the fixed points which holds approximatively on
length scales $L\gtrsim L_\Delta$. Eq. (\ref{flow2}) reduces then to a
single flow equation,
\begin{equation}
\label{flowD}
\frac{d\Delta}{dt}=\epsilon \Delta - \frac{9}{\pi^2}\Delta^2 
-\frac{p^6}{2}v^2\Delta^2.
\end{equation}
For finite $v$, the resulting phase diagram is characterized by a
critical line separating the flat and rough phases, along which the RG
flow is driven into a new critical $T=0$ fixed point with
$\Delta_F^*=\frac{\pi^2}{9}\epsilon(1-\mu)$ and
$v_F^*=\sqrt{\epsilon\mu/2}/p^2$ (see Fig. \ref{fig.dia}).
Linearizing around this fixed point, we obtain the critical exponents
of Eq.  (\ref{exponents}). This transition can be approached e.g. by
changing the temperature $T$ since $v^{1/2}/\Delta \sim
\rho_1^{1+p/2}$ where the CDW amplitude $\rho_1$ changes with $T$ and
vanishes at the Peierls temperature.

So far we have considered the case $\delta =0$. For non-zero $\delta$
the linear gradient term in (\ref{Ham}) can be integrated out leading
to $-\delta\gamma\int d^{d-1}{\bf y}\Delta\phi ({\bf y})$ where
$\Delta \phi ({\bf y})=\phi ({\bf y},L)-\phi ({\bf y},0)$.  In the C
phase $\Delta\phi ({\bf y})\equiv 0$ and hence the $\delta=0$
calculation applies. To estimate $\delta_c$, we compare in (\ref{Ham})
the term $\delta\partial\phi/\partial z$ with the cosine term on the
scale which corresponds to the renormalized soltion width (or
correlation length) $\xi$ leading to $\delta_c\approx
1/p\xi=gv^{1/2}(1-\Delta/\Delta_c)^\nu$.  In the I phase we rewrite
$\Delta\phi =\frac{2\pi L}{pl}$ and treat $l$ as a parameter to be
determined later by minimizing the free energy. With the
transformation $\phi ({\bf x})\rightarrow \phi ({\bf x})-\frac{2\pi
  z}{pl}$ we go back to periodic boundary conditions \cite{Ho83}
changing the cosine term to $g^2v\cos{(p\phi -\frac{2\pi z}{l})}$.  If
$\xi\ll l$ the RG up to $\xi$ is insensitive to the oszillations in
$z$ and the soliton lattice can be described in terms of an effective
free energy which includes also the steric repulsion between the walls
due to random field like disorder. The resulting domain wall density
vanishes as $l^{-1}\sim (\delta-\delta_c)^{\bar\beta}$ by approaching
the C phase where $\bar\beta=\zeta/2(1-\zeta)=(5-d)/2(d-2)$
\cite{Na83}
 follows
from the roughness exponent $\zeta=(5-d)/3$ \cite{GM82}.  Moreover,
the random potential destroys the translational long range order of
the soliton lattice, which shows an algebraic decay of correlations 
beyond $\tilde\xi_{\perp}\sim l$ and $\tilde\xi_{\parallel}\sim
l^{1/\zeta}$,
\begin{equation}
\label{algedec}
K({\bf x})=\rho_1^2\cos{\frac{2\pi z}{pl}\left(\left(
\frac{{\bf x}_{\parallel}}{\tilde{\xi}_{\parallel}}\right)^2 + 
\left(\frac{z}{\tilde{\xi}_{\perp}}
\right)^2\right)^{-\epsilon\pi^2/18p^2}}
\end{equation}
In the opposite case $\xi \gg l$, $v$ averages to zero on scales
larger than $l$ for geometric reasons.  The correlations show a
crossover to a behavior which is given by (\ref{algedec}) with
$\tilde{\xi}_\bot$, $\tilde{\xi}_\| \rightarrow L_\Delta$ and
exponent $-\epsilon\pi^2/18$.


As has been discussed earlier \cite{Na90,Bla}, a small applied
external force of density $f_{\rm ex}=eE$ from an electric field
applied along the chain direction leads for $f_{\rm ex}\ll f_c$, to a
creep motion of the CDW with a creep velocity
\begin{equation}
\label{creep}
v(f_{\rm ex})\sim \exp\left[-\frac{E_c}{T}\left(\frac{f_c}
{f_{\rm ex}}\right)^\kappa\right].
\end{equation}
In the U phase, $\kappa=(d-2)/2$, $f_c=f_{c,\Delta}$ and $E_c\approx
\gamma L_\Delta^{d-2}$, whereas in the C phase $\kappa=d-1$,
$f_c=f_{c,v}\approx\gamma\xi^{-2}$ and $E_c\approx \gamma\xi^{d-2}$.
Because of the pronounced difference in the exponents $\kappa$,
measurements of the creep velocity should give clear indications about
which phase is present.


Recently measured I-V curves of the conductor o-TaS$_3$ at
temperatures below 1K can be fitted by (\ref{creep}) with $\kappa=1.5$
-- $2$ \cite{Za97}.  The experimentally observed tendency to larger
$\kappa$ for purer crystals confirms the above interpretation.  In
several materials, such as ${\rm K}_{0.3}{\rm MoO}_3$, the periods are
near $p=4$ commensurability at low temperatures. For this material one
obtains $\xi_0 \approx 10^{-6}$cm \cite{Grbook}. The typical
parameters $\gamma V_0\approx 10^{-2}$eV, $\rho_1=10^{-2}$ and $v_{\rm
  F}=10^7{\rm cm}/{\rm sec}$ yield, after proper rescaling of
anisotropy, the estimate $L_\Delta\approx 10^{-4}$cm for an impurity
density of $100$ppm.  Thus it should be possible to see
commensurability effects if $\delta$ becomes small enough.


So far, we have excluded topological defects. These can be considered
if we treat $\phi({\bf x})$ as a multivalued field which may jump by
$\pm 2\pi$ at surfaces along which electrons are inserted to / omitted
from the chains at the intersection points of the chains and surfaces.
These surfaces are bounded by vortex lines. For $\delta=0=v$ it has
been argued recently that for weak enough disorder strength
$\Delta<\Delta_D$, the system is stable with respect to the formation
of vortices \cite{Gi95,GiKiFi}.  However, vortex lines will
proliferate for $\Delta>\Delta_D$. At present it is not clear whether
the corresponding transition is continuous or first order. For
$\delta=0$ but $v>0$ we expect that this transition extends to a line
$\Delta_D(v)$ until $v$ reaches a critical value $v_D$ with
$\Delta_D(v_D)=\Delta_c(v_D)$ (see Fig.  \ref{fig.dia}). For larger
$v$ the transition is probably in the universality class of the
$p$-state clock model in a random field, which has an upper critical
dimension $d_c=6$. In the Ising case $p=2$ the transition is of second
order \cite{Vi85}. A non-zero value of $\delta$ will in general
increase the size of the I phase, as schematically scetched in Fig.
\ref{fig.dia}b,c.


Our results can also be applied to the pinning of flux line lattices
(FLL) in type-II superconductors. In layered superconductors, the
CuO$_2$ planes provide a strong pinning potential favoring, for flux
lines oriented {\it parallel} to the layers, a smectic phase with
translational order present only along the layering axis \cite{Bal95}.
The influence of disorder on this phase is described by the CDW model
studied in this paper, if the CDW phase $\phi({\bf x})$ is identified
with the deviations of the smectic layers from their locked-in state.
Also a FLL oriented {\it perpendicular} to the layers in general feels
a {\em very} weak periodic potential of the underlying crystal, but
now the flux line displacements are described by a vector field. Since
also in this case weak disorder leads only to logarithmically growing
displacement of the flux lines \cite{Gi95}, a disorder driven
roughening transition of the CDW type studied above can be expected
for the FLL.  Furthermore, numerical simulations of FLL's are usually
modeled on a discrete lattice which acts for the FLL like a periodic
pinning potential. Thus the used grid size has to be small enough
(sufficiently large $p$) to avoid commensurability effects.


In conclusion, we have shown that a 3D CDW which is subject both to
impurity and lattice pinning undergoes a roughening transition from a
flat commensurate to a rough and in general incommensurate phase. On
the C side, this transition is described by three
independent critical exponents which obey the scaling relations for
random field systems with a discrete symmetry of the order parameter.
Compared to the thermal roughening transition in $d=2$, where the $T$
axis represents a line of fixed points, in the disorder driven case
the existence of one universal fixed point $\Delta^*_U$ leads to a
different RG flow. Notice that in the disorder dominated system
studied here, the important logarithmic roughness sets in only beyond
the Fukuyama-Lee length $L_\Delta$, whereas thermal roughness in $d=2$
appears on all length scales.  The changing creep behavior should be
an interesting consequence for the experimental observation of the
transition.

We acknowledge discussions with G.~Gr\"uner, P.~Monceau, V.L.~Pokrovsky,
S.~Scheidl and W.~Wonneberger.  We express our sincere gratitude to
L.~Balents who made numerous helpful suggestions.  This work was
supported by the SFB 341 and German-Israeli Foundation (GIF).

\end{multicols}{}

\begin{references}
\bibitem{Fu78b} After this paper was submitted for publication  
we learned that a one-dimensional version of the present model 
was considered already by H.~Fukuyama, J. Phys. Soc. Japan, {\bf 45}, 1474,
(1978), in which a similar estimate for the phase boundary was found.
\bibitem{subtleties} There may be
subtleties by exchanging the limits of large $L$ and small
$\epsilon$ which we neglected here, since the result is confirmed
by the RG presented below.
\end{references}
\end{document}